\documentclass[fleqn,usenatbib]{mnras}

\usepackage{newtxtext,newtxmath}
\usepackage[T1]{fontenc}

\DeclareRobustCommand{\VAN}[3]{#2}
\let\VANthebibliography\thebibliography
\def\thebibliography{\DeclareRobustCommand{\VAN}[3]{##3}\VANthebibliography}

\newcommand\z{3.98}
\newcommand\NH{21.44_{-0.33}^{+0.24}}
\newcommand\UH{-1.75_{-0.45}^{+0.28}}
\newcommand\Ne{2.6_{-0.7}^{+0.8}}
\newcommand\R{16_{-11}^{+23}}
\newcommand\Mdot{4100_{-2400}^{+6600}}
\newcommand\Edot{46.13_{-0.37}^{+0.41}}
\newcommand\Epercent{1.4_{-0.8}^{+2.2}}

\usepackage{graphicx}	% Including figure files
\usepackage{hhline}
\usepackage[flushleft]{threeparttable}
\usepackage{amsmath}

\title[BAL Outflow in SDSS J1130+0411]{High Mass Flow Rate in a BAL Outflow of Quasar SDSS J1130+0411}

\author[A. Walker et al.]{Andrew Walker$^{1}$, Nahum Arav$^{1}$, and Doyee Byun$^{1}$ \\
$^{1}$Department of Physics, Virginia Tech, Blacksburg, VA 24061, USA}

\date{Accepted 17 August 2022. Received 15 August 2022; in original form 12 May 2022}

\pubyear{2022}

\begin{document}
\label{firstpage}
\pagerange{\pageref{firstpage}--\pageref{lastpage}}
\maketitle

\begin{abstract}
We present the analysis of the absorption troughs of six outflows observed in quasar SDSS J1130+0411 ($z \approx \z$) with radial velocities ranging from $-2400$ to $-15,400$ km s$^{-1}$. These spectra were taken with the Very Large Telescope/Ultraviolet and Visual Echelle Spectrograph over the rest frame wavelength range of $1135-1890$ \AA. In the main outflow system ($v \approx -3200$ km s$^{-1}$), we identify \ion{Fe}{ii} and several \ion{Fe}{ii}* absorption troughs as well as \ion{Si}{ii} and \ion{Si}{ii}* troughs, which we use to determine the electron number density $\log n_e = \Ne$ cm$^{-3}$. Using the column densities of these and other ions, we determine a photoionization solution with hydrogen column density $\log N_H = \NH$ cm$^{-2}$ and ionization parameter $\log U_H = \UH$. From these values we derive the distance $R = \R$ kpc, the average mass flow rate $\dot{M} = \Mdot$ $M_{\odot}$ yr$^{-1}$, and the kinetic luminosity $\log \dot{E}_k = \Edot$ erg s$^{-1}$. This $\dot{E}_k$ is $\Epercent$\% of the quasar's Eddington luminosity, and therefore contributes significantly to AGN feedback.
\end{abstract}


\section{Introduction} \label{sec:intro}

Quasar outflows can be identified by absorption troughs that are blueshifted relative to the rest frame of active galactic nuclei (AGNs) (\textit{e.g.}, \citealt{Hewett_2003, Dai_2008, Knigge_2008}). Three main types of AGN absorption lines have been defined: broad absorption lines (BALs), narrow absorption lines (NALs), and mini-BALs. BALs are defined by continuous absorption below 90\% residual intensity with a velocity width $\Delta v \gtrsim 2000$ km s$^{-1}$ and are found in $\sim 20$\% of quasar spectra \citep{Hamann_2003, Stone_2019}. NALs have continuous absorption less than a few hundred km s$^{-1}$ wide, and are found in $\sim 60$\% of spectra \citep{Hamann_2003, Miller_2018, Stone_2019}. Mini-BALs are an intermediate category, with velocity widths $500 \lesssim \Delta v \lesssim 2000$ km s$^{-1}$ and a detection rate of $\sim 5$\% \citep{Hamann_2003,Rodriguez-Hidalgo_2012,Arav_2020_paper1}.

Quasar outflows can have a number of effects on the AGN environment, including contributing to the chemical evolution of their host galaxies by ejecting large quantities of metal and energy (\textit{e.g.}, \citealt{Di_Matteo_2005, Moll_2006}) and affecting star formation outside of the AGN by releasing angular momentum from accretion winds (\textit{e.g.}, \citealt{Murray_1995, Proga_2000}). Quasar outflows are also believed to have the potential to contribute to AGN feedback (\textit{e.g.}, \citealt{Ciotti_2009, Hopkins_2009, McCarthy_2010, Angles-Alcazar_2016, Vayner_2021}). Theoretical models suggest that for an outflow to effectively contribute to this feedback, its kinetic luminosity $\dot{E}_k$ must be at least $0.5\%-5\%$ of the Eddington luminosity ($L_\mathrm{Edd}$) of the quasar's supermassive black hole (SMBH) (\citealt{Hopkins_2009, Scannapieco_2004}, respectively). Observations show that these criteria can be met (\textit{e.g.}, \citealt{Moe_2009, Arav_2013_HE0238, Arav_2020_paper1, Chamberlain_2015, Xu_2019, Xu_2020a_paper2, Xu_2020b_paper6, Miller_2020a_paper3, Miller_2020c_paper7, Choi_2020, Byun_2022_J1439-0106, Byun_2022_J0242+0049, Choi_2022}). The distance between the central source and the outflow ($R$) and the total hydrogen column density ($N_H$) are crucial in determining $\dot{E}_k$, and thus whether the outflow could have a significant role in AGN feedback. We can determine $R$ by finding the ionization parameter ($U_H$) and the electron number density ($n_e$), as many studies in the past have done (\textit{e.g.}, \citealt{de_Kool_2001, de_Kool_2002, Hamann_2001, Gabel_2005, Borguet_2012, Xu_2018, Arav_2020_paper1, Miller_2020a_paper3, Byun_2022_J1439-0106, Byun_2022_J0242+0049, Choi_2022}). The column densities of various ions in a system can be used to find $N_H$ and $U_H$, and $n_e$ can be found from the ratios of excited state to ground state ionic column densities \citep{Arav_2018}. In this paper we make a determination of $R$, $N_H$, and $\dot{E}_k$ for an outflow found in the Very Large Telescope/Ultraviolet and Visual Echelle Spectrograph (VLT/UVES) spectrum of SDSS J113010.58+041128.0 (hereafter J1130+0411).

We obtained the data for this object from the Spectral Quasar Absorption Database (SQUAD) data release 1 \citep{Murphy_2018_squad}, a survey of 475 quasars whose spectral data were collected from VLT/UVES. While searching through this database, we found that J1130+0411 possessed prominent absorption, including excited state lines, which suggested an outflow. We derive a systemic redshift $z=\z$ (see Section \ref{sec:redshift}), and we identify the outflow velocity $v=-3200$ km s$^{-1}$ system to be BAL system, as the \ion{C}{iv} trough has $\Delta v \sim 2200$ km s$^{-1}$. Here we focus on this system as within it we observe excited states of \ion{Fe}{ii} and \ion{Si}{ii}. These lines are important as they are largely unblended in the spectrum, and so we can use them in order to determine $n_e$, and thus $R$. Additionally, we find that this BAL system has a particularly high average mass flow rate $\dot{M} = \Mdot$ $M_{\odot}$ yr$^{-1}$ that is, for example, almost ten times higher than the largest $\dot{M}$ objects seen in the \cite{Choi_2022} FeLoBAL sample and comparable to an outflow system in J0242+0049 analyzed by \cite{Byun_2022_J1439-0106} (see Section \ref{sec:comparison}).

This paper is organized as follows. In Section \ref{sec:obs}, we present the details of the observation and data acquisition of J1130+0411. Section \ref{sec:analysis} describes the analysis process including the spectral fitting to measure ionic column densities and the determination of the parameters $N_H$, $U_H$, and $n_e$. In Section \ref{sec:energy} we then utilize these values and present our calculations of $R$, $\dot{M}$, and $\dot{E}_k$. Section \ref{sec:dis} discusses our results, compares with other works, and describes the other seven absorption components in J1130+0411. Section \ref{sec:sum} summarizes this paper. We adopt a cosmology of $h=0.696$, $\Omega_m=0.286$, and $\Omega_\Lambda=0.714$ \citep{Bennett_2014}.

\section{Observations} \label{sec:obs}

\begin{figure*}
    \centering
    \includegraphics[width=0.99\linewidth]{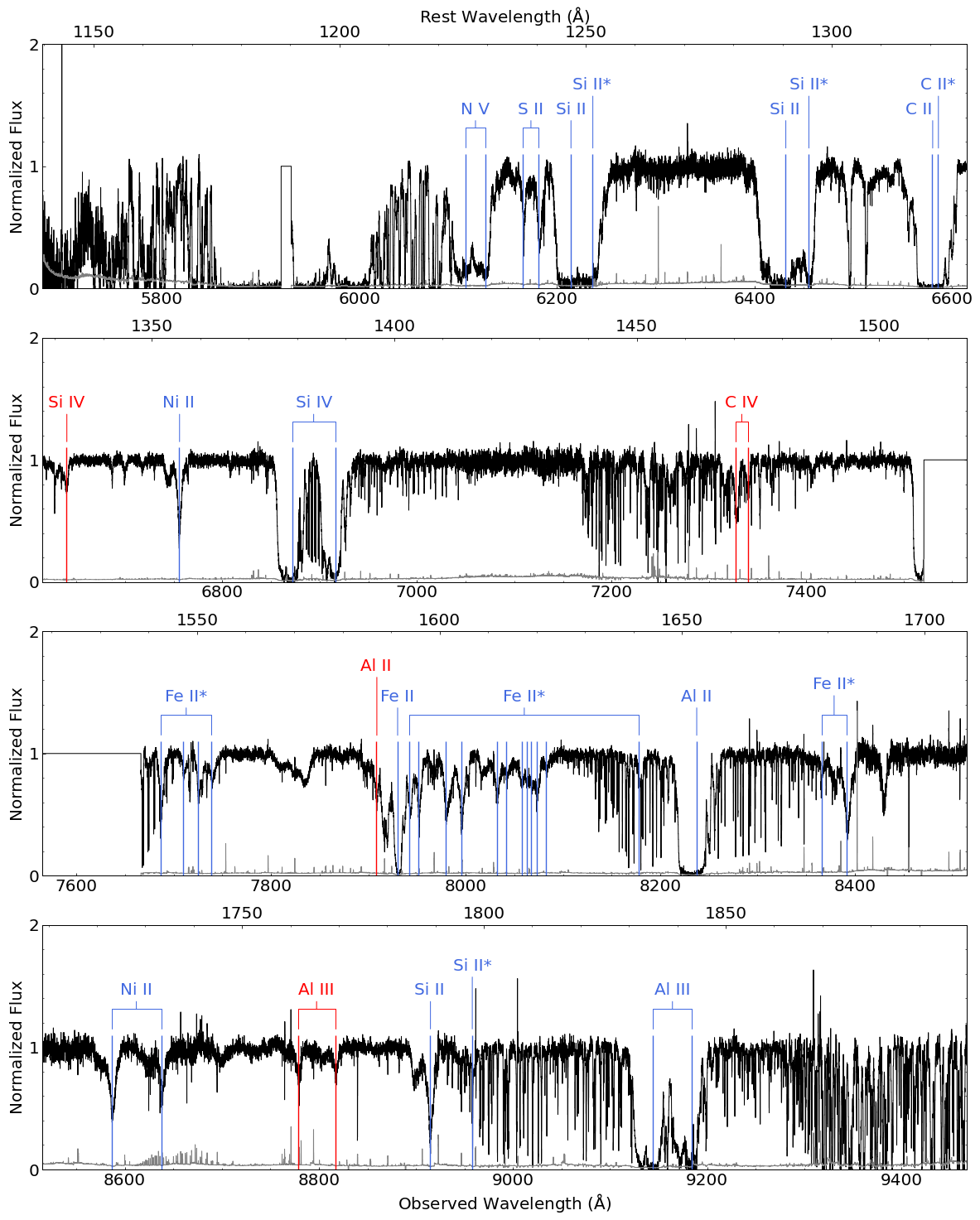}
    \caption{Normalized spectrum of J1130+0411 from the 2018 SQUAD data release \citep{Murphy_2018_squad}. Troughs of the main system S5 are presented in blue. Troughs of the highest velocity system S1, identified by \protect\cite{Chen_2021_CIVcatalog}, are presented in red. Only the red absorption trough of \ion{Si}{iv} in S1 is labeled, as the blue trough is blended with the \ion{C}{ii} troughs of lower velocity systems and we do not obtain a column density from it. Note that one of the \ion{Si}{ii}/\ion{Si}{ii}* BALs ($1527$ \AA \ \& $1533$ \AA \ rest wavelength) and the \ion{C}{iv} BAL are in a wavelength gap not covered by UVES ($7521-7666$ \AA \ observed wavelength).}
    \label{fig:uvesspec}
\end{figure*}

The quasar J1130+0411 (J2000: RA=11:30:10, DEC=+04:11:28, $z=\z$, see Section \ref{sec:redshift}) was observed on February 12 and 14, 2008 with VLT/UVES as part of the program 080.B-0445(A), with overall wavelength coverage $3670-9467$ \AA. The spectral data has a resolution $R \simeq 40,000$ and a signal-to-noise ratio $S/N \simeq 37$. The data was combined, reduced, and normalized by \cite{Murphy_2018_squad} as part of their SQUAD Data Release 1. Figure \ref{fig:uvesspec} displays the full normalized spectrum. We identify eight absorption systems in this quasar, named S1-8 in order of decreasing velocity (see Table \ref{tab:othersystems}). The main focus of this paper is on the BAL outflow, S5, while the other systems are covered in Section \ref{sec:other}.

\setlength{\tabcolsep}{12pt}
\renewcommand{\arraystretch}{1.4}

\begin{table*}
    \centering
    \caption{Properties of All Absorption Systems in J1130+0411}
    \begin{threeparttable}
    \begin{tabular}{c c c c c c}
    \hline
        System     & Velocity    & log($N_H$)        & log($U_H$)     & Status  \\
                   & (km s$^{-1}$)      & (cm$^{-2}$)  &     & \\
    \hline
        S1   & $-15,400$      & ...      & ...  & outflow           \\
        S2   & ...\tnote{a}      & ...      & ...  & intervening       \\  
        S3   & ...\tnote{b}       & $21.10_{-2.80}^{+0.65}$      & $-1.47_{-2.37}^{+0.62}$  & intervening       \\  
        S4   & $-3600$       & $21.79_{-2.48}^{+0.01}$      & $-1.18_{-2.13}^{+0.03}$  & S5 subcomponent      \\
        S5   & $-3200$       & $\NH$      & $\UH$  & outflow (main)           \\
        S6   & $-2700$       & $>19.95_{-1.04}$      & $-3.17_{-0.12}^{+0.58}$  & outflow           \\  
        S7   & $-2600$       & ...      & ...  & outflow  \\  
        S8   & $-2400$       & $>19.97_{-0.13}$      & $>-2.99_{-1.03}$  & outflow  \\  
    \hline
    \end{tabular}
    \begin{tablenotes}
    \item[a] The redshift of the intervening system S2 is $z=3.798$.
    \item[b] The redshift of the intervening system S3 is $z=3.912$.
    \end{tablenotes}
    \end{threeparttable}
    \label{tab:othersystems}
\end{table*}

\subsection{Determination of Redshift} \label{sec:redshift}
\begin{figure*}
    \centering
    \includegraphics[width=0.95\linewidth]{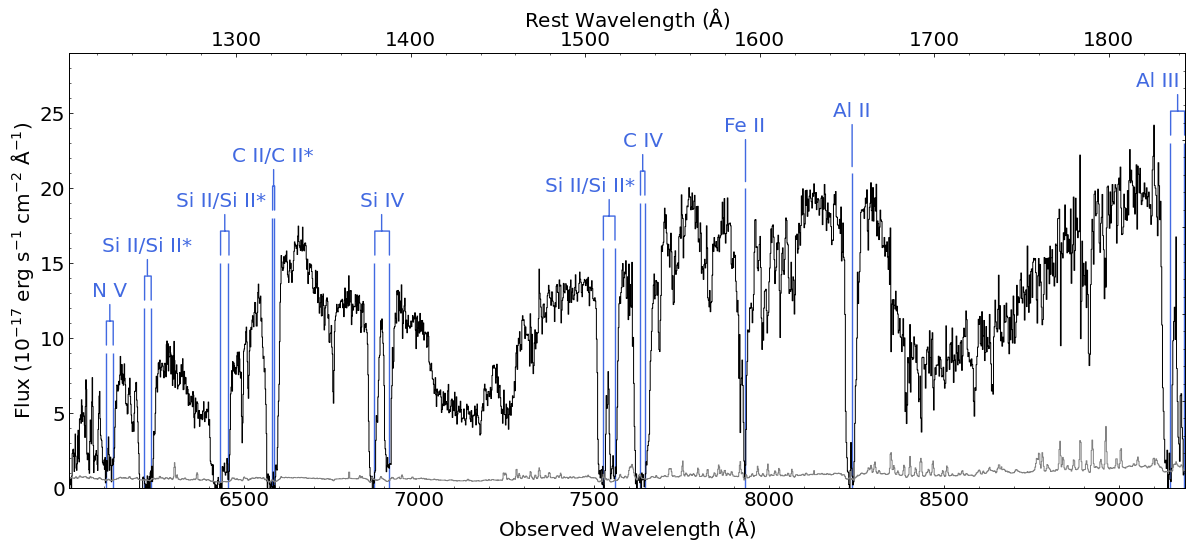}
    \caption{Un-normalized spectrum from SDSS. The locations of prominent BALs are marked in blue.}
    \label{fig:sdssspec}
\end{figure*}

In order to measure the width of the \ion{C}{iv} BAL, measure the width of the \ion{C}{iv} emission line (see Section \ref{sec:feedback}), and determine the redshift of J1130+0411, we retrieved an SDSS spectrum from MJD=52642 (January 3, 2003). Figure \ref{fig:sdssspec} shows this unnormalized spectrum with the main absorption troughs labeled. Note that the vertical lines are not centered in many of the BALs. The high column density of these ions allows for higher velocities in the outflow, thus widening the blue side of the BAL. Additionally, S4 is blended with the blue side of these BALs.

For J1130+0411 we determine a systemic redshift $z=\z$ that lines up with the \ion{C}{ii}/\ion{C}{ii}* and \ion{Si}{iv} emission lines. As an additional check, this redshift places the Ly $\alpha$ line $\sim 6$ \AA \ blueward of the edge of the Lyman $\alpha$ forest in the rest frame. This estimate is different from the one made by \cite{Chen_2021_CIVcatalog}, who determine a redshift of $z=3.930$ that appears to instead line up with the BAL outflow.  \cite{Chen_2021_CIVcatalog} also identify one \ion{C}{iv} absorption system. Using their reported redshift to calculate absorption system velocities, we find that this system best matches our highest velocity system, S1. They did not identify S5, as this \ion{C}{iv} absorption falls in the spectral gap of this UVES data ($7521-7666$ \AA \ observed wavelength).

Our redshift is also different from the redshift reported on the SDSS website, $z=5.25$. SDSS uses an algorithm in order to determine the redshift of objects in its catalog, and this algorithm does not always produce accurate measurements. It is clear by eye that their reported redshift does not coincide with any emission features in the unnormalized spectrum.

\section{Data Analysis} \label{sec:analysis}

\subsection{Spectral Fitting} \label{sec:fit}

The first step in determining the physical characteristics of an outflow is to determine the column densities of the ions observed in the system. The simplest way of doing this is to use the apparent optical depth (AOD) method, which assumes that the outflow completely and uniformly covers the source \citep{Savage_1991}. In this case, the normalized flux is related to the optical depth by
\begin{equation}
    I(v) = \mathrm{exp}(-\tau(v)),
\end{equation}
where $I$ is the observed flux and $\tau$ is the optical depth. The column density can found by
\begin{equation}
    N_\mathrm{ion} = \frac{m_e c}{\pi e^2 f \lambda} \int \tau(v) dv,
\end{equation}
where $m_e$ is the electron mass, $c$ is the speed of light, $e$ is the elementary charge, and $f$ and $\lambda$ are the oscillator strength and wavelength of the transition line, respectively. This method only gives a lower limit to the column density. If an ion presents as a doublet or multiplet, we can use the more accurate partial covering (PC) method (\citealt{Barlow_1997, Arav_1999b, Arav_1999a}), which accounts for effects such as non-black saturation (\citealt{Edmonds_2011, Borguet_2012}). This method introduces a velocity-dependent covering fraction $C(v)$ (\citealt{de_Kool_2002, Arav_2005}), and the normalized flux depends on the optical depth by
\begin{align}
\begin{split}
    I_1(v) & = [1-C(v)] + C(v)e^{-\tau(v)}, \\
    I_2(v) & = [1-C(v)] + C(v)e^{-R_{21}\tau(v)},
\end{split}    
\end{align}
where $I_1$ and $I_2$ are the normalized fluxes of the doublet transition lines, and $R_{21} = f_2\lambda_2/f_1\lambda_1$. Here, $f_1$, $f_2$, $\lambda_1$, and $\lambda_2$ are the oscillator strengths and wavelengths of the two doublet lines. We adopt column densities calculated using PC rather than AOD whenever possible.

Many of the absorption lines are blended with other troughs, so we model each trough by fitting to an unblended section of the trough a Gaussian of the form
\begin{equation}
    \tau_i(v) = \frac{A_i}{\sigma_i \sqrt{2\pi}}*\mathrm{exp} \left(\frac{-(v-v_i)^2}{2\sigma_i^2} \right),
\end{equation}
\begin{equation}
    I_i(v) = \mathrm{exp}(-\tau_i(v))
    \label{equ:gaussfitflux}
\end{equation}
where, for trough $i$, $A_i$ is the scaling factor, $\sigma_i$ is the velocity dispersion ($\mathrm{FWHM}=2\sigma \sqrt{2\mathrm{ln}(2)}$), and $v_i$ is the velocity centroid. For example, in many cases the blue wing of the trough is the only unblended portion and therefore is what the Gaussian is fitted to. For S5 in particular, we observe both a wide and a narrow component in most lines, so for this system we use a double Gaussian to fit the troughs, \textit{i.e.}, $I_i=I_\mathrm{wide} \times I_\mathrm{narrow}$.

Compared with \ion{Si}{ii} $1260$ \AA \ and $1304$ \AA, the \ion{Si}{ii} $1808$ \AA \ trough is shallow and narrow, and so we assume it is unsaturated and interpret its AOD column density as a measurement rather than a lower limit. We observe non-black saturation in the \ion{Al}{iii} $\lambda\lambda1854,1860$ BAL, so we obtain a measurement for its column density using the PC method. For all other BALs, we use the AOD method to obtain a lower limit. Table \ref{tab:columndensities} gives the column densities we adopted in this analysis. We add 20\% error in quadrature to the column density errors in order to account for uncertainty in the continuum model \citep{Xu_2018}. Note that the adopted \ion{Fe}{ii} column density is somewhat higher than that of the AOD value. This is because four of the \ion{Fe}{ii}* energy levels show more than one trough for the same level, and so for these we use the PC solution. We then add these measurements to the AOD measurements of the remaining levels in order to calculate the total \ion{Fe}{ii} column density.

\renewcommand{\arraystretch}{1.4}

\begin{table}
    \centering
    \caption{Column Densities of Ions in Main System of J1130+0411 (in units of $10^{12}$ cm$^{-2}$)}
    \begin{tabular}{l c c c}
    \hline
        Ion     & AOD  & PC  &   Adopted      \\
    \hline
        \ion{C}{ii}      & $6700_{-26}^{+950}$  & ...  & $>6700_{-1300}$           \\
        \ion{N}{v}      & $2900_{-21}^{+23}$  & ...  & $>2900_{-580}$            \\
        \ion{Al}{ii}     & $180_{-1}^{+25}$  & ...  & $>180_{-40}$              \\
        \ion{Al}{iii}     & $780_{-15}^{+82}$  & $860_{-140}^{+10}$  & $860_{-220}^{+170}$        \\
        \ion{Si}{ii}     & $31000_{-1500}^{+1500}$  & ...  & $31000_{-6400}^{+6400}$   \\
        \ion{Si}{iv}     & $1100_{-10}^{+13}$  & ...  & $>1100_{-220}$            \\   
        \ion{S}{ii}      & $350_{-16}^{+16}$  & $720_{-140}^{+120}$  & $720_{-200}^{+190}$       \\ 
        \ion{Fe}{ii}     & $11000_{-950}^{+950}$  & ...  & $>18000_{-2700}$          \\ 
        \ion{Ni}{ii}     & $880_{-66}^{+66}$  & $900_{-230}^{+230}$  & $900_{-290}^{+290}$             \\
    \hline
    \end{tabular}
    \label{tab:columndensities}
\end{table}

\subsection{Photoionization Modelling} \label{sec:nvu}
\begin{figure}
    \centering
    \includegraphics[width=0.97\linewidth]{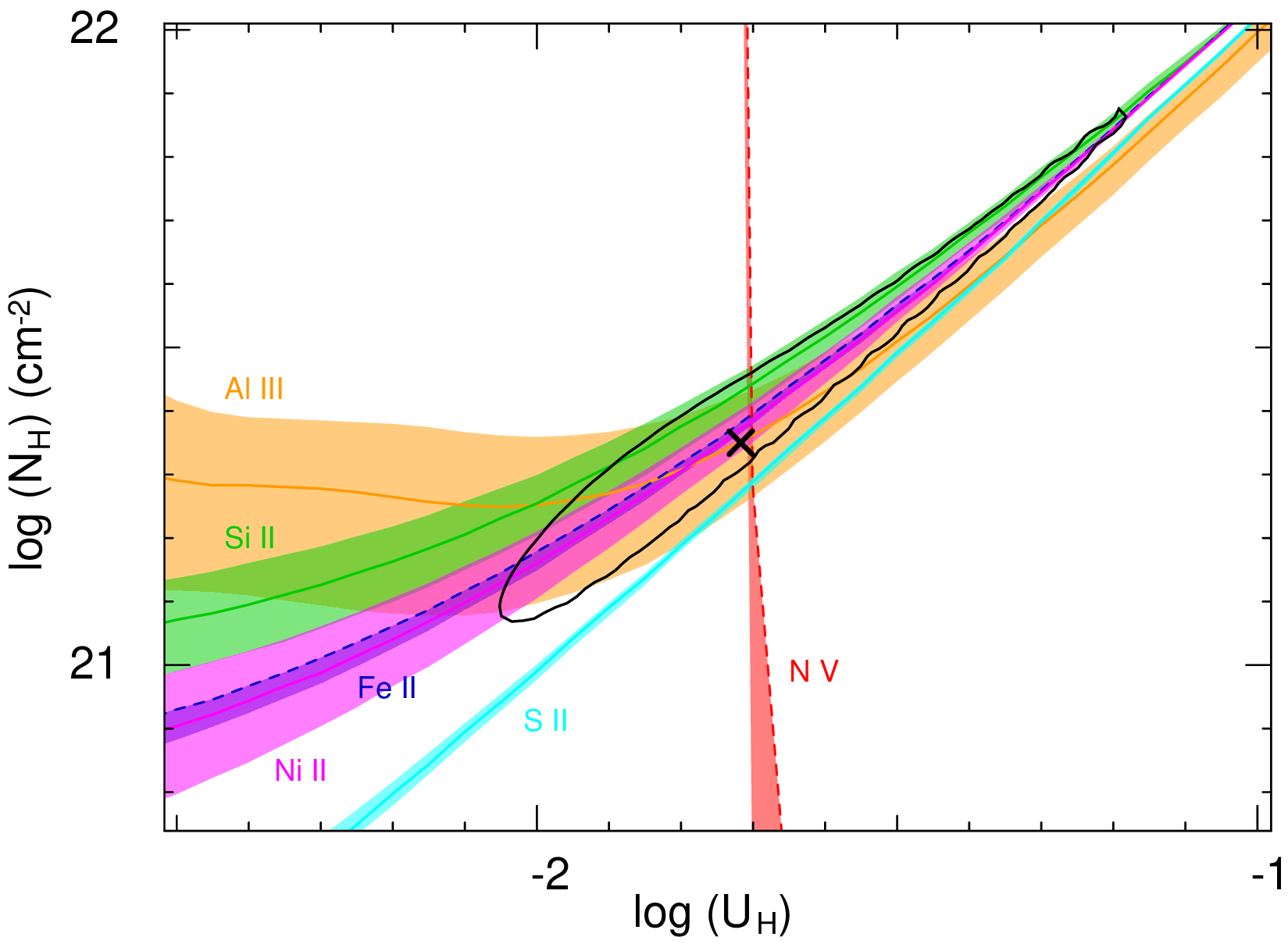} %Change width to 0.99 for two column (0.5 for one column)
    \caption{Photoionization solution for the main system of J1130. The colored lines represent the allowed values of $N_H$ and $U_H$ for a given $N_\mathrm{ion}$. Solid lines represent measurements, and dashed lines represent lower limits. The colored ribbons around these lines represent the errors of these measurements. The solution, found by $\chi^2$ minimization, is displayed as a black cross, and the black contour represents the $1\sigma$ error for this solution. Only the ions that affected the solution are shown.}
    \label{fig:photosolution}
\end{figure}

In order to determine $N_H$ and $U_H$, we use the spectral synthesis code Cloudy (version c17.00, \citealt{Ferland_2017_cloudy}) to create photoionization models with a given hydrogen column density ($N_H$) and ionization parameter ($U_H$), following previous works (\textit{e.g.}, \citealt{Xu_2019, Miller_2020a_paper3, Miller_2020c_paper7, Byun_2022_J1439-0106, Byun_2022_J0242+0049}). For these models we assume solar metallicity and the spectral energy distribution (SED) of quasar HE0238-1904 (\citealt{Arav_2013_HE0238}, see their Figure 10). Together, these parameters determine the column density of each ion in a photoionization model. We can therefore create a grid of models over a range of $N_H$ and $U_H$ and compare them to our measured ionic column density. The model with the lowest $\chi^2$ gives a solution that best matches our observations. The best-fit solution for S5 gives us $\log N_H = 21.34_{-0.33}^{+0.24}$ cm$^{-2}$ and $\log U_H = \UH$. The "Adopted" column of Table \ref{tab:columndensities} lists the column densities used for this fit. Figure \ref{fig:photosolution} shows a visualization of this solution.

\subsection{Electron Number Density} \label{sec:sss}

The ratios of excited state to ground state column densities can be used to determine the electron number density of an absorber \citep{Moe_2009}. This can become difficult to measure directly if the ground or excited state troughs in a spectrum are significantly blended, or if the ground state trough is saturated. In this case, we can use the Synthetic Spectral Simulation method (SSS) in order to help determine $n_e$ \citep{Xu_2020a_paper2}. For this method, we input the previously obtained photoionization solution, with a range of $n_e$, into Cloudy to get the total column densities of the desired ions. From the wavelength, oscillator strength, excited ionic critical densities, and output column density of each absorption line, we create synthetic, theoretical spectra with the range of $n_e$. We then find the synthetic spectrum that best matches the observed spectrum in order to determine the best fit $n_e$ for that absorption system. Since we use a fixed metallicity and SED, $N_H$ affects the total strength of all lines, and $U_H$ affects the relative strength between different ions. Excited ions are mostly populated by collisions between free electrons and ground state ions \citep{Osterbrock_2006_agn2}, so $n_e$ affects the relative strength between the ground state and various excited states of ions.

\begin{figure*}
    \includegraphics[width=0.49\linewidth]{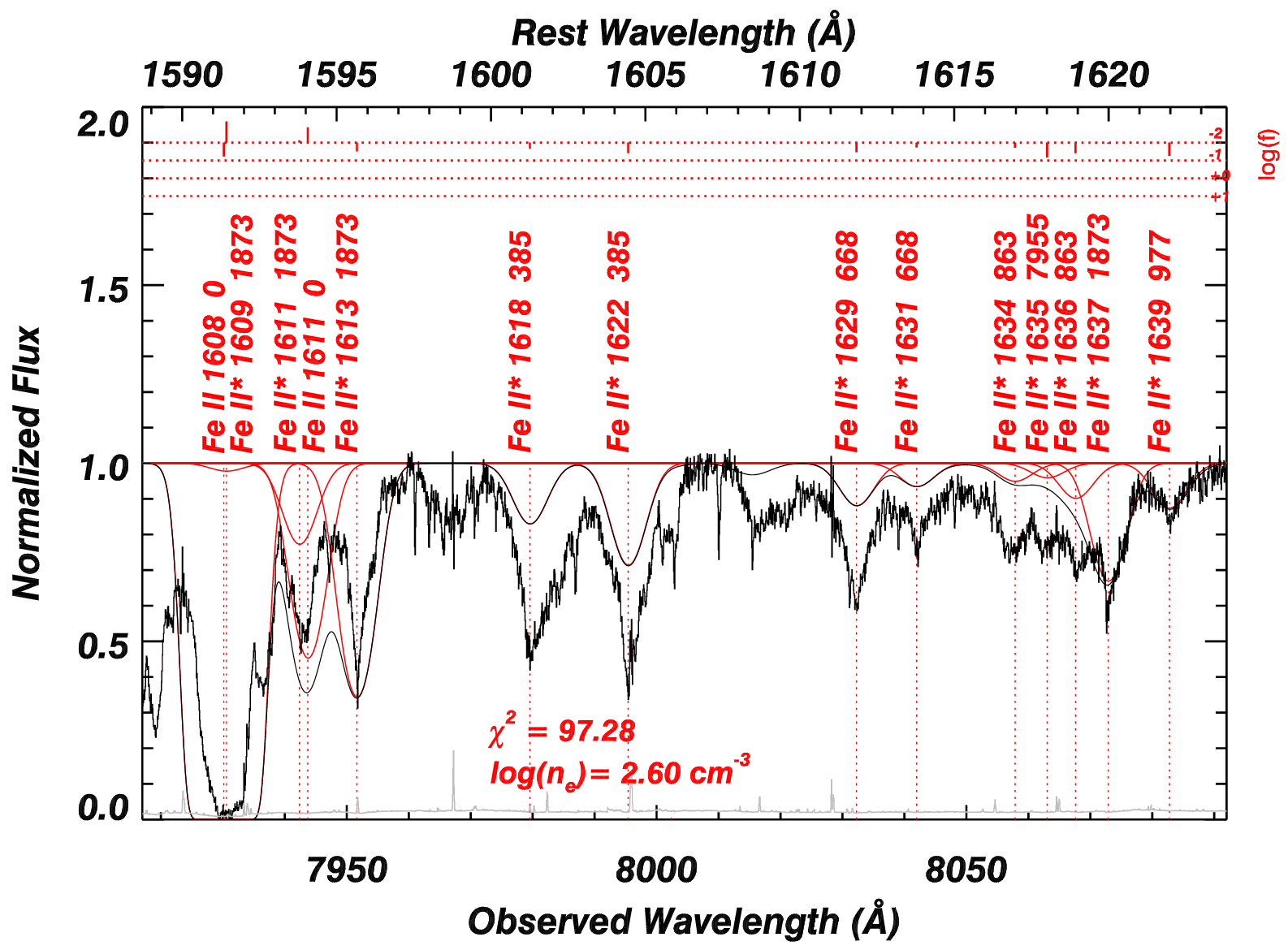}
    \includegraphics[width=0.49\linewidth]{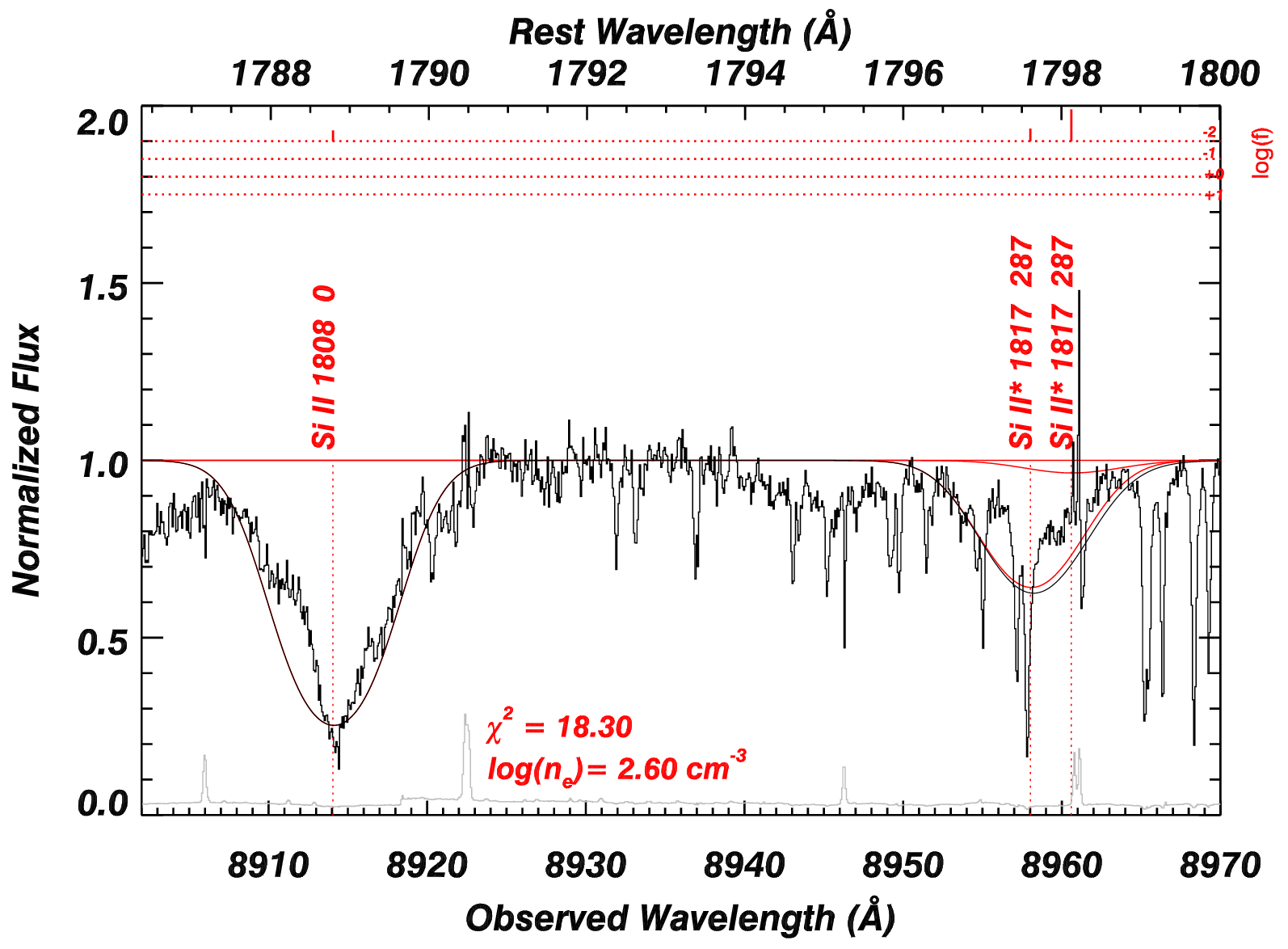}
    \caption{SSS plots of the \ion{Fe}{ii}/\ion{Fe}{ii}* and \ion{Si}{ii}/\ion{Si}{ii}* lines in the main system of J1130+0411 that we use to find $n_e$. The logarithm of the oscillator strength of each line is shown in red at the top with a separate y axis. The $n_e$ used to create the spectrum is shown at the bottom of the plot, along with the reduced $\chi^2$.}
    \label{fig:xinfengsss}
\end{figure*}

Table \ref{tab:properties} gives a determination of $n_e$ for S5 of J1130+0411 and Figure \ref{fig:xinfengsss} shows the SSS plots used to find this solution. Note that these fits have a high reduced $\chi^2$ (also given in Figure \ref{fig:xinfengsss}). Certain phenomena unrelated to fitting $n_e$ contribute to this, such as the unidentified line at observed wavelength $7968$ \AA \ in the left plot of Figure \ref{fig:xinfengsss} and the atmospheric lines from $8932-8970$ \AA \ in the right plot. However, the largest contribution to this $\chi^2$ comes from the fact that the best fit SSS method overpredicts the strength of the \ion{Fe}{ii}* $1873 \ \mathrm{cm}^{-1}$ energy level transitions and underpredicts the strength of all other \ion{Fe}{ii}* energy level transitions, resulting in a poor fit overall. This discrepancy is discussed further in Section \ref{sec:comparison}. In order to determine the positive and negative errors, we increase and decrease (respectively) the input $n_e$ until the simulated spectrum has double the $\chi^2$ compared with the best fit. From the SSS method we determine $\log n_e = \Ne$ cm$^{-3}$. We also note that increasing $N_H$ by $0.1$ dex slightly increases the quality of the fit, so we adopt for this system $\log N_H = \NH$ cm$^{-2}$.

\section{Distance \& Energetics} \label{sec:energy}

We can determine the distance $R$ from the outflow to the central source using the definition of the ionization parameter,
\begin{equation}
    U_H \equiv \frac{Q_H}{4\pi R^2 n_H c},
\end{equation}
where $Q_H$ is the incidence rate of hydrogen-ionizing photons, $n_H$ is the number density of hydrogen, which for a highly ionized plasma is estimated as $n_e \approx 1.2n_H$, and $c$ is the speed of light \citep{Osterbrock_2006_agn2}.

We can determine $Q_H$ by taking the continuum flux at rest wavelength $\lambda = 1495$ \AA \ from the SDSS data ($F_\lambda = 1.32_{-0.07}^{+0.07} \times 10^{-16}$ erg s$^{-1}$ cm$^{-2}$ \AA$^{-1}$) and scaling it to match the HE0238 SED. Integrating this spectrum over all energies above $1$ Ryd gives $Q_H = 5.8_{-0.3}^{+0.3} \times 10^{57}$ s$^{-1}$, and the corresponding bolometric luminosity is $L_\mathrm{bol} = 9.2_{-0.5}^{+0.5} \times 10^{47}$ erg s$^{-1}$. For this system we determine $R=\R$ kpc.

With the distance of the outflow, we can calculate the outflow mass \citep{Borguet_2012},
\begin{equation}
    M \simeq 4\pi \Omega R^2 N_H \mu m_p,
\end{equation}
where $\Omega$ is the global covering factor (\emph{i.e.}, the fraction of the total solid angle $4\pi$ of the quasar that the outflow covers, $\sim0.2$ for BALQSOs; \citealt{Hewett_2003}; see Section 5.2 of \citealt{Dunn_2010}), $\mu=1.4$ is the mean atomic mass per proton, and $m_p$ is the proton mass. From there, we can divide by the dynamical time scale to get the average mass flow rate,
\begin{equation}
    \dot{M} \simeq 4\pi \Omega R N_H \mu m_p v,
\end{equation}
where $v$ is the outflow velocity. The kinetic luminosity is then
\begin{equation}
    \dot{E}_k \simeq \frac{1}{2}\dot{M}v^2.
\end{equation}
For this system we calculate $\dot{M} = \Mdot$ $M_{\odot}$ yr$^{-1}$ and $\log \dot{E}_k = \Edot$ erg s$^{-1}$. Table \ref{tab:properties} summarizes all calculated properties.

%$\log \dot{E}_k = \Edot\, {\rm erg~s^{-1}}$

\subsection{Error Propagation}

The thin, elliptical shape of the photoionization solution (see Figure \ref{fig:photosolution}) demonstrates that the $N_H$ and $U_H$ errors are correlated with each other. In order to see how this could affect our error estimates, we calculate three values $\zeta^+$, $\zeta$, and $\zeta^-$ that correspond to the top-right corner, middle solution, and bottom-left corner of the photoionization solution contour, respectively, so that $\dot{M}=\zeta*\sqrt{Q_H/n_e}$:
\begin{align}
\begin{split}
    &\zeta^+ = \mu m_p 4\pi \Omega v*(N_H + \sigma_{N_H}^+)  *\sqrt{1.2/4 \pi c *(U_H + \sigma_{U_H}^+)}, \\
    &\zeta = \mu m_p 4\pi \Omega v*N_H  *\sqrt{1.2/4 \pi c *U_H}, \\
    &\zeta^- = \mu m_p 4\pi \Omega v*(N_H - \sigma_{N_H}^-)  *\sqrt{1.2/4 \pi c *(U_H - \sigma_{U_H}^-)},
\end{split}    
\end{align}
where $\sigma_{N_H}^+$, $\sigma_{N_H}^-$, $\sigma_{U_H}^+$, and $\sigma_{U_H}^-$ are the positive and negative errors for $N_H$ and $U_H$ given in Table \ref{tab:properties}. Now, we want to make sure that $\zeta^+ > \zeta$ and $\zeta^- < \zeta$. Since $\zeta \propto N_H/\sqrt{U_H}$, this requires satisfying the following condition: $$\frac{N_H}{\sqrt{U_H}} < \frac{(N_H+\sigma_{N_H}^+)}{\sqrt{U_H+\sigma_{U_H}^+}}, \hspace{1cm} \frac{N_H}{\sqrt{U_H}} > \frac{(N_H-\sigma_{N_H}^-)}{\sqrt{U_H-\sigma_{U_H}^-}}.$$ If these inequalities do not hold, then we simply swap the definitions of $\zeta^+$ and $\zeta^-$. Finally, we define $\sigma_{\zeta}^+ = \zeta^+ - \zeta$ and $\sigma_{\zeta}^- = \zeta - \zeta^-$ and add to this in quadrature the errors of $n_e$ and $Q_H$:
\begin{equation}
    \log \sigma_{\dot{M}} = \sqrt{\left(\log \sigma_{\zeta}\right)^2+\frac{1}{4}\left[\left( \log \sigma_{n_e} \right)^2+\left(\log \sigma_{Q_H} \right)^2 \right]},
\end{equation}
where the factor of $1/4$ comes from the fact that $R$ is proportional to the square root of $n_e$ and $Q_H$. In the end, for S5, this method produces errors that are a factor of $\sim 1.3$ lower than those calculated by simply adding the $R$ and $N_H$ errors in quadrature.

\setlength{\tabcolsep}{12pt}
\renewcommand{\arraystretch}{1.3}

\begin{table}
    \centering
    \caption{Properties of the Main System in J1130+0411}
    \begin{tabular}{r l}
    \hline
        log ($N_H$)     & $\NH$   \\[-1.2mm]  (cm$^{-2}$)     &   \\
        log ($U_H$)     & $\UH$   \\[-1.2mm]  (dex)           &   \\
        log($n_e$)      & $\Ne$       \\[-1.2mm]  (cm$^{-3}$)     &   \\
        $R$             & $\R$   \\[-1.2mm]  (kpc)            &   \\
        $\dot{M}$       & $\Mdot$      \\[-1.2mm]  ($M_{\odot}$ yr$^{-1}$)     &   \\
        log ($\dot{E}_k$)       & $\Edot$   \\[-1.2mm]  (erg s$^{-1}$)      &   \\   
        $\dot{E}_k/L_{\mathrm{Edd}}$    & $\Epercent$    \\[-1.2mm]  (\%)    &   \\
    \hline
    \end{tabular}
    \label{tab:properties}
\end{table}

\section{Discussion} \label{sec:dis}

\subsection{Contribution to AGN Feedback} \label{sec:feedback}

\begin{figure}
    \centering
    \includegraphics[width=0.99\linewidth]{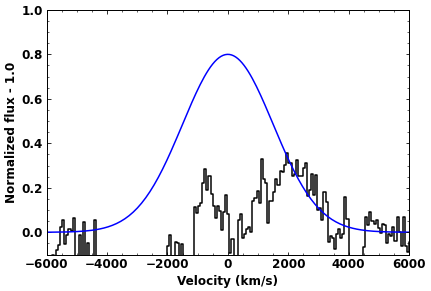} %Change width to 0.99 for two column (0.5 for one column)
    \caption{Gaussian fit to the C IV emission. Normalized flux - 1 as a function of velocity is displayed in black. The Gaussian curve is displayed in blue, and is centered on $v=0$ km s$^{-1}$. Note that there is heavy absorption in the vicinity of \ion{C}{iv} $\lambda1549$. We fit the Gaussian by eye to the relatively unblended right wing of the \ion{C}{iv} emission.}
    \label{fig:CIVemission}
\end{figure}

In order to numerically compare our calculated $\dot{E}_k$ with $L_\mathrm{Edd}$, we must first determine $L_\mathrm{Edd}$ for J1130+0411. To do this, we take advantage of a method introduced by \cite{Coatman_2017}, in which they use the width of the \ion{C}{iv} emission line to determine the mass of the black hole (see their Equations 4 \& 6). This method is based on that used by \cite{Vestergaard_2006}. Modelling the C IV emission in this object is difficult as it is heavily blended with absorption. However, it is the only emission feature visible in the spectrum of J1130+0411 that we can use to determine the black hole mass. So, to approach this we center a Gaussian on the \ion{C}{iv} $\lambda 1549$ \AA \ emission line and fit it to an unblended portion of the red wing in order to get the full width at half maximum (FWHM). Figure \ref{fig:CIVemission} shows this Gaussian fit. From there it is simple to calculate $L_\mathrm{Edd}$ \citep{Osterbrock_2006_agn2}, and we derive for this system $L_\mathrm{Edd} = 9.7_{-1.4}^{+1.7} \times 10^{47}$ erg s$^{-1}$. Our calculated $\dot{E}_k$ for S5 is then $\Epercent$\% of the $L_\mathrm{Edd}$ of J1130+0411. This is above the $0.5$\% threshold suggested by \cite{Hopkins_2009}, thus we conclude that the main system has the potential to contribute to AGN feedback.

\subsection{Comparison with Other Outflows} \label{sec:comparison}

Other observations of outflows containing \ion{Fe}{ii}* absorption have been performed. In SDSS J1439-0106, \cite{Byun_2022_J1439-0106} compare the abundance ratios of five different \ion{Fe}{ii}* energy levels to the ground state and find that the $1873 \ \mathrm{cm}^{-1}$ level is in agreement with the other four. \cite{Xu_2021_Fe} measure multiple excited \ion{Fe}{ii} energy levels up to $21,580 \ \mathrm{cm}^{-1}$ in Q0059-2735, however no $1873 \ \mathrm{cm}^{-1}$ troughs were reported. \cite{Choi_2020} also report multiple excited states with the exclusion of $1873 \ \mathrm{cm}^{-1}$ in SDSS J1352+4239. Several excited levels are measured in QSO 2359-1241 by \cite{Korista_2008}, including three $1873 \ \mathrm{cm}^{-1}$ lines: $2332$ \AA, $2349$ \AA, and $2361$ \AA. However as they note, these troughs are blended with other lines. All of these lines are out of range of the J1130+0411 data. It appears that the \ion{Fe}{ii} $2332$ \AA, $2349$ \AA, $2361$ \AA, and $2494$ \AA \ lines detected in FBQS 0840+3633 by \cite{de_Kool_2002} are of the $1873 \ \mathrm{cm}^{-1}$ excited level, but these are also heavily blended and out of the spectral range of J1130+0411.

Blending of \ion{Fe}{ii}* $1873 \ \mathrm{cm}^{-1}$ with other \ion{Fe}{ii} lines is also a problem in J1130+0411. Although this is somewhat mitigated by using the SSS method, this method is not perfect, and assumes an AOD scenario for every trough. The spectral synthesis procedure \textit{SimBAL} \citep{Leighly_2018_FeLoBAL} provides a similar approach to SSS, but instead implements a power law method \citep{Arav_2005, Leighly_2019_FeLoBAL} in order to create synthetic spectra. Future studies could use \textit{SimBAL} to attempt to model the \ion{Fe}{ii} lines in J1130+0411 in that way.

We also note that the average mass flow rate of $\dot{M} = \Mdot$ $M_{\odot}$ yr$^{-1}$ is particularly large. \cite{Byun_2022_J0242+0049} find an even higher mass flow rate of $6500_{-3400}^{+8900}$ $M_{\odot}$ yr$^{-1}$ in their S2 of quasar SDSS J0242+0049. These rates are roughly ten times greater than the highest of which reported by \cite{Choi_2022}, with the largest in their sample, J1154+0300, having only $\dot{M} \sim 500$ $M_{\odot}$ yr$^{-1}$. Comparing J1130+0411 with relatively similar objects in this sample, we conclude that our larger $\dot{M}$ can be attributed to both a higher outflow velocity and outflow distance in S5. Our $L_\mathrm{bol} = 9.2_{-0.5}^{+0.5} \times 10^{47}$ erg s$^{-1}$ is higher than the largest bolemetric luminosity in the \cite{Choi_2022} sample by $\sim 0.5$ dex, and our $\log \dot{E}_k = \Edot$ is comparable to the the highest values in their sample. Additionally of note, the systemic redshift $z=\z$ of J1130+0411 is significantly higher than the lower redshift quasars, with $0.66 < z < 1.63$, targeted by \cite{Choi_2022}.

\subsection{Other Absorption Systems in J1130+0411} \label{sec:other}

We identify seven absorption systems in this quasar other than the main system. We order these in terms of decreasing velocity, so that S1 ($-15,400$ km s$^{-1}$) is the highest velocity system and S8 ($-2400$ km s$^{-1}$) is the lowest velocity system. S1 was identified by \cite{Chen_2021_CIVcatalog} (see Section \ref{sec:redshift}). The presence of wide, smooth \ion{C}{iv} troughs and the unblended red \ion{Si}{iv} trough lead us to believe that this system is an outflow. S2 has \ion{C}{iv} and \ion{Si}{iv} troughs, but as they are very narrow, this is most likely an intervening system. Although we observe \ion{C}{ii}* in S3, the fact that we do not observe any \ion{Si}{iv} is again indicative of an intervening system. Many of the expected lines in S4 fall in the blue end of the S5 BALs, leading us to believe that this system is actually subcomponent of S5. S5 is the main system discussed in the bulk of this paper. In S6, we see some excited state ions, though blended, and many of the ions we expect to see in quasar outflows, so this system is most likely an outflow. S7 and S8 also show (blended) \ion{C}{ii}* troughs, so these are most likely outflows as well. 

Unfortunately, all of these systems aside from S5 have few unblended diagnostic troughs from which to extract reliable ionic column densities. In addition, no unblended excited troughs are identified in any of these weaker systems. Therefore, while we are able to get some information about $N_H$ and $U_H$ for some systems, we are not able to extract any electron number densities, and so we are not able to calculate any distances or energetics for these systems. Table \ref{tab:othersystems} summarizes the properties of these systems that we were able to measure. Since S2 and S3 are intervening and thus their redshifts are likely cosmological, these redshifts are reported rather than their velocities in the footnotes of this table.

\section{Summary} \label{sec:sum}

We have presented the analysis of eight absorption systems, which we label S1-8, in quasar SDSS J1130+0411, with a focus on S5 in particular. The data for this object was taken by VLT/UVES and was pulled from the first SQUAD data release \citep{Murphy_2018_squad}. We determine a new systemic redshift $z = \z$ that differs both from the one reported by \cite{Chen_2021_CIVcatalog} and that by the SDSS catalog. By measuring the column densities of the nine ions given in Table \ref{tab:columndensities}, we found a photoionization solution for $N_H$ and $U_H$. We also determined information about $N_H$ and $U_H$ for some of the other systems using photoionization analysis.

By comparing the trough depth of excited and ground states of \ion{Fe}{ii} and \ion{Si}{ii}, we used the SSS method \citep{Xu_2020a_paper2} so as to determine a best-fit $n_e$. However, this best-fit synthetic spectrum was a rather poor fit, as it overpredicts the \ion{Fe}{ii}* $1873 \ \mathrm{cm}^{-1}$ troughs while underpredicting all other observed \ion{Fe}{ii}* troughs. Additional observations of \ion{Fe}{ii}* excited states including the $1873 \ \mathrm{cm}^{-1}$ excited state will show whether this is a recurring phenomenon in quasar spectra.

With measurements of $N_H$, $U_H$, and $n_e$, we determine an outflow distance $R = \R$ kpc and a kinetic luminosity to Eddington luminosity ratio $\dot{E}_k/L_\mathrm{Edd} = \Epercent$\%. Since models suggest that $\dot{E}_k$ must be at least $0.5\%-5\%$ of $L_\mathrm{Edd}$ (\citealt{Hopkins_2009, Scannapieco_2004}, respectively), we conclude that this system is capable of contributing to AGN feedback.

\section*{Acknowledgements}
NA, AW, and DB acknowledge support from NSF grant AST 2106249, as well as NASA STScI grants AR-15786, AR-16600 and AR-16601. We also thank the anonymous referee whose constructive comments and suggestions helped to clarify and improve this paper.

\section*{Data Availability}
The reduced and normalized VLT/UVES data for J1130+0411 is a part of the first SQUAD data release presented in \cite{Murphy_2018_squad}. The SDSS data used to determine the systemic redshift, the width of the \ion{C}{iv} emission, and the width of the \ion{C}{iv} BAL is available on the SDSS archive.

\bibliographystyle{mnras}
\bibliography{_paper}

\bsp	% typesetting comment
\label{lastpage}
\end{document}